\pdftrailerid{} 
\documentclass[conference]{IEEEtran}

\usepackage{graphicx}
\usepackage{subfig}
\usepackage[T1]{fontenc}
\usepackage[scaled=0.9]{DejaVuSansMono}
\usepackage{listings}
\usepackage{textcomp}
\usepackage{multirow}
\usepackage{array}
\usepackage{cite}
\usepackage{url}
\usepackage{balance}
\usepackage{amsmath}
\usepackage{algorithm2e}
\usepackage{float}
\usepackage{xcolor}
\usepackage{pdfprivacy}
\usepackage{fancyvrb}

\newcolumntype{L}[1]{>{\raggedright\let\newline\\\arraybackslash\hspace{0pt}}m{#1}}
\newcolumntype{C}[1]{>{\centering\let\newline\\\arraybackslash\hspace{0pt}}m{#1}}
\newcolumntype{R}[1]{>{\raggedleft\let\newline\\\arraybackslash\hspace{0pt}}m{#1}}

\lstnewenvironment{code}[1][]%
  {
    \noindent
    \minipage{\linewidth}
    \medskip
    \lstset{basicstyle=\ttfamily\footnotesize,frame=single,#1}}
  {\endminipage}

\makeatletter
\def\footnoterule{\relax%
  \kern-5pt
  \hbox to \columnwidth{\hfill\vrule width 0.5\columnwidth height 0.4pt\hfill}
  \kern4.6pt}
\makeatother

\graphicspath{{images/}}
\DeclareGraphicsExtensions{.pdf,.jpeg,.png,.jpg,.eps}

\renewcommand{\IEEEkeywordsname}{Keywords}

\begin{document}
\title{Hypersparse Traffic Matrix Construction using GraphBLAS on a DPU  }

\author{\IEEEauthorblockN{
William Bergeron$^1$, Michael Jones$^1$, Chase Barber$^2$, Kale DeYoung$^2$, George Amariucai$^2$,  \\ Kaleb Ernst$^2$, Nathan Fleming$^2$, Peter Michaleas$^1$, Sandeep Pisharody$^1$, \\ Nathan Wells$^2$, Antonio Rosa$^1$, Eugene Vasserman$^2$, Jeremy Kepner$^1$}
\IEEEauthorblockA{$^1$MIT, $^2$KSU}
}

\maketitle

\IEEEtitleabstractindextext{
\begin{abstract}
Low-power small form factor data processing units (DPUs) enable  offloading and acceleration of a broad range of  networking and security services. DPUs have accelerated the transition to programmable networking by enabling the replacement of FPGAs/ASICs in a wide range of network oriented devices. The GraphBLAS sparse matrix graph open standard math library is well-suited for constructing anonymized hypersparse traffic matrices of network traffic which can enable a wide range of network analytics.  This paper measures the performance of the GraphBLAS on an ARM based NVIDIA DPU (BlueField 2) and, to the best of our knowledge, represents the first reported GraphBLAS results on a DPU and/or ARM based system.  Anonymized hypersparse traffic matrices were constructed at a rate of over 18 million packets per second. 

\end{abstract}
}
\let\thefootnote\relax\footnotetext{This material is based upon work supported by the Under Secretary of Defense for Research and Engineering under Air Force Contract No. FA8702-15-D-0001. Research was also sponsored by the United States Air Force Research Laboratory and the Department of the Air Force Artificial Intelligence Accelerator and was accomplished under Cooperative Agreement Number FA8750-19-2-1000. The views and conclusions contained in this document are those of the authors and should not be interpreted as representing the official policies, either expressed or implied, of the Under Secretary of Defense for Research and Engineering, Department of the Air Force, or the U.S. Government. The U.S. Government is authorized to reproduce and distribute reprints for Government purposes notwithstanding any copyright notation herein.}
\IEEEpeerreviewmaketitle
\IEEEdisplaynontitleabstractindextext
\section{Introduction}
\label{sec:introduction}
Data processing units (DPUs), like NVIDIA's BlueField 2, allow for data centers and supercomputers to offload, accelerate, and isolate a broad range of advanced networking, security services, and other infrastructure functions and control-plane applications \cite{burstein2021nvidia}.  Utilizing an energy efficient onboard ARM CPU to offload  tasks onto edge devices can significantly reduce server power consumption.  We explore using a DPU to construct anonymized hypersparse traffic matrices for an edge network device traffic.  GraphBLAS is ideally suited for both constructing and analyzing anonymized hypersparse traffic matrices \cite{kepner2016mathematical,jones2022graphblas}.  The performance of the SuiteSparse GraphBLAS library \cite{davis2019algorithm} on an NVIDIA DPU  is demonstrated and the performance for varying numbers of processes and threads is measured.  This performance demonstrates that anonymized hypersparse traffic matrices are readily computable on edge network devices with minimal compute resources. 

\section{Technology}
\label{sec:current}
Data processing units (DPUs) have emerged as a new computing pillar in an ever-evolving computing landscape, joining  central processing units (CPUs) and graphics processing units (GPUs).  A DPU is a system-on-a-chip combining a standard programmable multi-core CPU, such as an  ARM processor, with a high-performance network interface capable of efficiently transferring data to the host's CPU and/or GPU.  The DPU market has advanced rapidly and vendors now include NVIDIA, Marvell, Fungible/Microsoft, Broadcom, Intel, and AMD Pensando.  
    
    
    \begin{figure}[ht]
\center{\includegraphics[width=1\columnwidth]{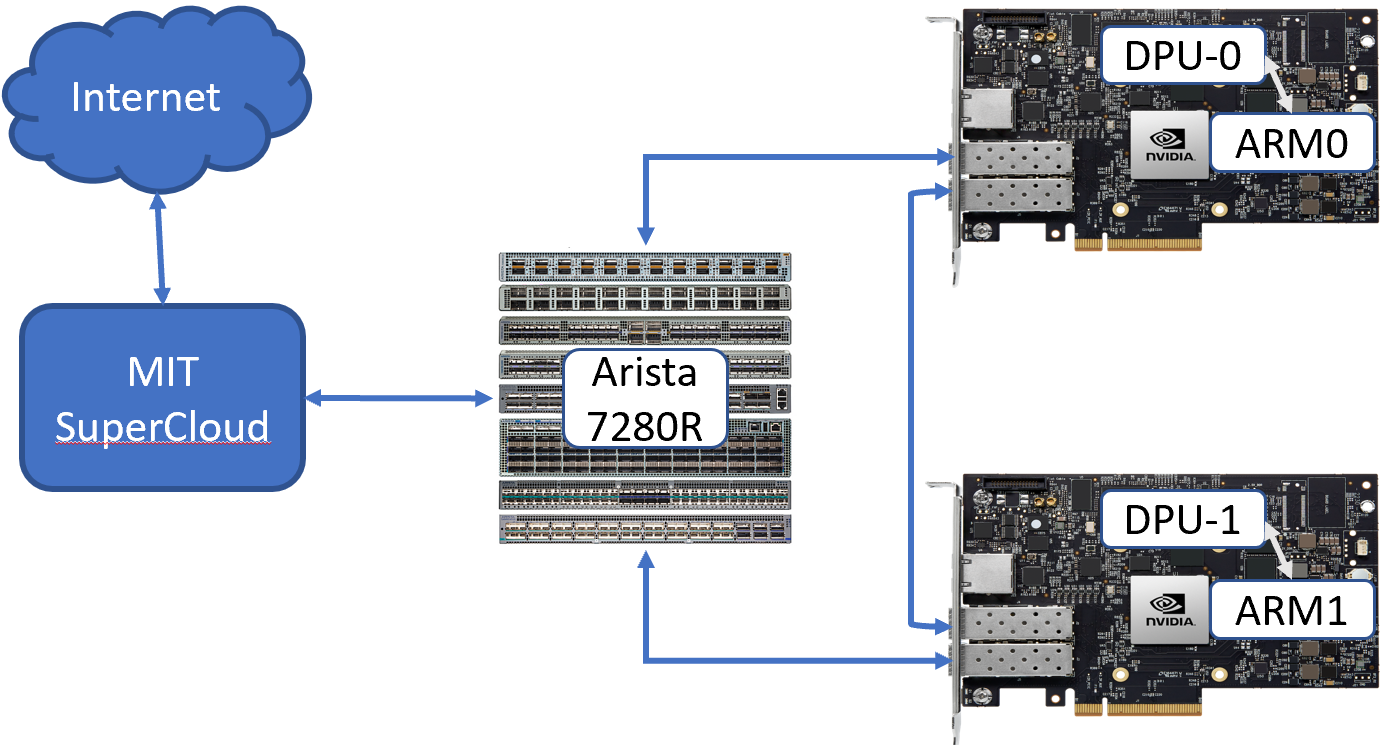}}
	\caption{\textit{DPU Hardware Configuration on MIT SuperCloud}}
	\label{fig:capture}
\end{figure} 
The hardware used for the network data collection and GraphBLAS matrix creation were two NVIDIA BlueField-2 DPUs. These Mellanox based PCI devices have a variety  of accelerated software-defined networking, storage, security, and management services. The NVIDIA DOCA (Data center On-a-Chip Architecture) software framework enables developers to create  applications and services for NVIDIA DPUs.
GraphBLAS is an API specification that defines the standard construction for graph algorithms using linear algebra.  GraphBLAS utilizes sparse matrix construction to represent graphs as either an adjacency matrix or an incidence matrix. Graph operations, matrix traversal and matrix transformation are implemented via linear algebraic methods like matrix multiplication over different semirings defined in the GraphBLAS specification. GraphBLAS is an ongoing community effort, including representatives from industry, academia, and government research labs \cite{kepner2016mathematical,bulucc2017design,davis2019algorithm,jones2022graphblas}.
        

\section{Implementation}
\label{sec:implimentation}
Two NVIDIA Mellanox MT42822 BlueField-2s each with 8 ARMv8 A72 cores\cite{MT42822} were installed in separate compute nodes belonging to the the MIT SuperCloud, and both were connected to a single Arista DCS-7280 switch on an isolated VLAN at 10 Gbit so that traffic could be sent between the two devices.[Fig. \ref{fig:capture}]  This configuration allowed for the scalable generation of network traffic between the DPUs via two methods: the application dpdk-burst-replay in conjunction with a supplied packet capture (PCAP) file, and Intel's pktgen, a DPDK-based traffic generator able to send wire-rate traffic using 64-byte frames.

As traffic passed through the DPU, each inbound packet was handled by the network interface belonging to the embedded ARM processor.
The DPDK Ethernet device was configured to create and bind a specified number of receive queues so network traffic could be initialized when run.  A hardware flow rule was installed to limit the processing to Ethernet packets. The flow rules determined how packets got flagged, dropped, and queued during run-time. The DPU implementation collected packets via a capture loop from the flows at run-time.

The SuiteSparse GraphBLAS library was used to create $2^{32}{\times}2^{32}$ hypersparse network traffic matrix ${\bf A}$, using the source and destination IP addresses in the IP headers of the received Ethernet packets as row and column identifiers, with the matrix value ${\bf A}(i,j)$ containing the number of packets sent from source $i$ to destination $j$.  This matrix provides a useful map of the incoming network traffic that can be use for a variety of analytics \cite{trigg2022hypersparse}.


Two modes were tested: GraphBLAS only and GraphBLAS+IO.  In the GraphBLAS only mode, 8 batches of 64 traffic windows each containing $2^{17}$ random source/destination pairs were generated.  A hypersparse $2^{32}{\times}2^{32}$ GraphBLAS  traffic matrix was constructed for each traffic window and timed.  This GraphBLAS only mode was repeated for 1, 2, 4, and 8 concurrent instances of the program corresponding to using all 8 ARM cores on the DPU.  In the GraphBLAS+IO mode, pairs of threads were executed on one DPU that received simulated random packets from the other DPU.  One thread received the packets and the other thread constructed the hypersparse GraphBLAS  traffic matrix in manner similar to the GraphBLAS only mode.  This GraphBLAS+IO mode executed using  2, 4, and 8 concurrent threads corresponding to using all 8 ARM cores on the receiving DPU.

\section{Results}
\label{sec:results}

\begin{figure}
\center{\includegraphics[width=1\columnwidth]{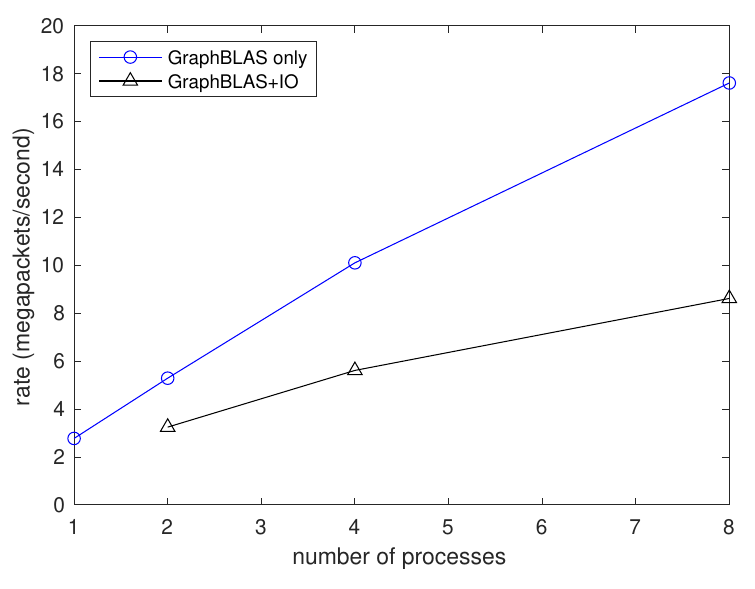}}
	\caption{\textit{Average GraphBLAS only and GraphBLAS+IO rates}}
	\label{fig:Matrix-creation}
\end{figure}

The packet rates for the GraphBLAS only and GraphBLAS+IO modes are are shown in Figure \ref{fig:Matrix-creation}.  Both modes show good scaling.  The GrapBLAS only mode peaked at 18 million packets per second, while the GraphBLAS+IO mode peaked at 8 million packets per second.

GraphBLAS can use OpenMP for multi-threaded processing.  OpenMP was tested in the GraphBLAS only case and the small size of the GraphBLAS matrices ($2^{17}$ entries) provided insufficient work to see benefits from OpenMP.

\section{Conclusion}
\label{sec:conclusion}
The performance of the GraphBLAS on an ARM based NVIDIA DPU (BlueField 2) had been measured.  Anonymized hypersparse traffic matrices were constructed at a rate of over 18 million packets per second, which is comparable to a 200 Gigabit network link (assuming 10,000 bits/packet).  The results demonstrate the viability of GraphBLAS on these types of edge devices. To the best of our knowledge, this represents the first reported GraphBLAS results on a DPU and/or ARM based system.  
 

\section*{Acknowledgments}
\label{sec:acknowledgments}
The authors wish to acknowledge the following individuals: W. Arcand, S. Atkins, D. Bestor, C. Birardi, B. Bond, A. Bowne, S. Buckley, C. Byun,  K Claffy, C. Conrad, Chris Demchak, A. Edelman, E. Ferber, G. Floyd, V. Gadepally, J. Gottschalk, O. Green, D. Gupta, C. Hill, M. Houle, A. Klien, C. Leiserson, A. Levanon, K. Malvey,  S. Mashhadi, J. McDonald, C. Milner, S. Mohindra, L. Milechin, J. Mullen, R. Patel, S. Pentland, H. Perry, C. Prothmann, A. Prout, A. S. Rejto, Reuther, J. Rountree, D. Rus, S. Samsi, R. Shah,  M. Sherman, S. Weed, C. Yee, M. Zissman.


\bibliographystyle{unsrt}
\bibliography{DPU}

\end{document}